\documentclass[aps,11pt]{revtex4}
\parskip=\medskipamount

\newcommand{\be}{\begin{equation}}
\newcommand{\ee}{\end{equation}}

\usepackage{amsmath,amssymb}

\DeclareMathOperator*{\Tr}{Tr}

\def \beq {\begin{equation}}
\def \eeq {\end{equation}}
\def \beqn {\begin{eqnarray}}
\def \eeqn {\end{eqnarray}}

\renewcommand\Im{\operatorname{Im}}

\def \l {\left(}
\def \r {\right)}

\def \lal {\langle}
\def \rr {\rangle}

\def \ep {\epsilon}

\global\long\def\p{\partial}

\begin{document}
\title{The Toda system and solution to the $\cal{N}$=2 SUSY Yang-Mills theory}

\author{A. Gorsky$^{1,2}$}

\affiliation{
$^1$Institute for Information Transmission Problems of the Russian Academy of Sciences, Moscow,
Russia, \\ $^2$Moscow Institute of Physics and Technology, Dolgoprudny 141700, Russia 
}

\begin{abstract}
We briefly review the place of  the Toda closed chain and the Toda field theory  in solution to the 
$\cal{N}$=2 supersymmetric Yang-Mills
(SYM) theory. The classical and quantum aspects of the correspondence are mentioned and the role
of branes as  degrees of freedom is emphasized.
\end{abstract}

\maketitle

\section{Introduction}

The family of Hamiltonian systems invented by Toda 50 years ago \cite{toda} serves in theoretical physics as the sample model for 
integrable finite-dimensional systems and 2d field theories in many occasions. Sometimes it provides the idealized system to be perturbed 
in some way to fit the problem under consideration, sometimes it describes the physical systems exactly when it has enough symmetry. The Toda system is uniquely
attributed to the group-like phase spaces. For  the open finite-dimensional Toda chains the finite-dimensional groups
are relevant while the affine groups provide the phase space for the closed Toda chain. The Toda field theories also can be 
considered as the simple system at the Kac-Moody related phase spaces. The classical and quantum aspects of the Toda systems are
reviewed in \cite{olsha}.

The classical closed finite-dimensional Toda chain has appeared in the interesting way in Seiberg-Witten solution
to the 4D $\cal{N}$=2 pure SYM theory \cite{SW1,SW2}. It has been recognized for the $SU(2)$ SYM theory  in \cite{gkmmm}
and the guess about the solution  of $SU(N)$ SYM theory was based at correspondence with
$SU(N)$ closed Toda chain \cite{MW}. Later two more integrable systems which have Toda closed chain 
as a proper limit were identified in SYM theory with some additional matter. Namely the $\cal{N}$=$2^*$ $SU(N)$ SYM with adjoint 
matter corresponds to the $SU(N$) elliptic Calogero model \cite{dw} while the $SU(N)$ SYM with 
fundamental matter \cite{chain,ggm1} to the particular spin chains. The limit to the closed Toda chain
corresponds to the decoupling of the heavy matter in the asymptotically free theory when the scale
in the pure YM theory emerges from the dimensional transmutation. If the 4D SYM gets lifted to the 
$(4+1)$ dimensional theory with one compact dimension the closed Toda chains gets lifted to the
relativistic closed Toda chain \cite{nikita}. The review of the early development of these issues can be found 
in \cite{gmir}.

In spite of the great effectiveness of the integrability approach for derivation 
of low-energy effective actions for SYM the origin of the Toda-like 
classical Hamiltonian systems and their generalizations for the theories with matter remained obscure 
for a while. In particular it was
not clear precisely what are degrees of freedom in these systems and why these systems are not
quantized. Two major achievements allowed to fill these gaps in understanding. First, the
introduction of the $\Omega$-background allowed to derive the Seiberg-Witten solution
microscopically from the instanton ensemble \cite{nekrasov,NO02}. Later it was shown  \cite{NS1,NS2,NS3} that in the particular 
limit of $\Omega$-background $\epsilon_1=0, \epsilon_2 \neq 0$ the integrable system gets
quantized and the Planck constant gets identified with $\epsilon_2$. 

Secondly the AGT 
correspondence  between the $\cal{N}$=2 SYM theories and Liuoville
for $SU(2)$ SYM and Toda field theories for $SU(N)$ SYM  has been formulated \cite{agt,wyllard}. 
The Nekrasov instanton partition functions were identified with the particular conformal blocks 
in Liouville and Toda field theories.
One can add into the corresponding conformal block the additional operator degenerated
at the second level which induces the second order differential equation of the
Schrodinger type for the conformal block. It is solution to this equation for the conformal block with insertion  
of degenerate operator 
coincides with the wave function of the finite-dimensional Hamiltonian system. That is the coordinates of the
Toda chain or Calogero model are the positions of the full surface defects in the internal space. 
The very approach providing the derivation of the conformal block and therefore the Nekrasov
partition function  via the insertion of the
additional degenerate operator is the particular realization of the Zamolodchikov's monodromy
method \cite{zamo}. This method allows to evaluate the conformal block itself from the monodromy
of the solutions to the  decoupling  equation.

We will very briefly review in the simple terms the ideas and steps which have lead to the current 
understanding of the place for the Toda integrable systems  in the rich world
of $\cal{N}$=2 SUSY YM systems.

\section{Summary on the Seiberg-Witten solution}

The Seiberg-Witten solution to the $\cal{N}$=2 SYM theory \cite{SW1,SW2} provides
the answer for the low-energy effective action and the spectrum of the stable
BPS particles. To get the low-energy effective action it is necessary to take into 
account a one-loop contribution and the sum up the infinite instanton series.
The latter problem looked hopeless however the combination of the deep
physical guesses and some mathematical tools provided the answer without
the microscopic evaluation of the instanton sum.

Let us describe several steps which lead to the explicit solution \cite{SW1,SW2}

\begin{itemize}
\item {Holomorphy}. 

The Lagrangian  involves the
potential term for the scalar fields $\phi$  of the form
\be
V(\phi)=\Tr[\phi,\phi^{+}]^{2},
\ee
where the trace is taken over adjoint representation of the gauge group
$SU(N)$. It gives rise to the valleys in the theory, when
$[\phi,\phi^+]=0$.
The vacuum energy vanishes along the valleys, hence,
the supersymmetry remains unbroken. One may always choose the v.e.v.'s of the
scalar field to
lie in the Cartan subalgebra $\phi=diag(a_{1},...,a_{n})$. These parameters
$a_i$ can not serve, however, good order parameters, since there is still
a residual Weyl symmetry which changes $a_i$ but leave the same vacuum state
and one should consider the set of the gauge invariant
order parameters $u_{k}=<\Tr\phi^{k}>$ that fix the vacuum state unambiguously.
Hence we obtain a moduli space parametrized
by the vacuum expectation values of the scalars, which is known
as the Coulomb branch of the whole moduli space of the theory.

The choice of the point on the Coulomb branch is
equivalent to the  choice of the vacuum state and simultaneously it
yields the scale which the coupling is frozen on. At the generic point of
the moduli space, the theory becomes effectively Abelian after the
condensation of the scalar.
As soon as the scalar field acquires the vacuum expectation value, the
standard
Higgs mechanism works and there emerge heavy gauge bosons at large values
of the vacuum condensate. 

The initial action of $\cal{N}$=2  theory written in $\cal{N}$=2  superfield $\Psi$  has the simple
structure 
\beq 
S(\Psi )=\Im\tau \int \Tr\Psi^{2} \qquad
\tau=\frac{4\pi i}{g^{2}} + {\theta\over 2\pi}
\eeq
where g is coupling constant and $\theta$ - the standard coefficient
in front of topological charge. 
The $\cal{N}$=2 supersymmetry
implies
that the low energy effective action gets renormalized only by holomorphic
contributions so that it is ultimately given by a single function
known as prepotential $S_{eff}(\Psi )=\Im\int{\cal{F}}(\Psi)$. The
prepotential
is a holomorphic function of moduli $u_k$ 
except for possible singular points at the values of moduli where additional
massless states can appear disturbing the low energy behavior.

Thus, the problem effectively reduces to the determination of single
holomorphic function. If one manage to fix its behavior nearby singularities,
the function can be unambiguously restored. One of the singularities,
corresponding
to large values of v.e.v.'s, i.e. to the perturbative limit is under control. 
All other singularities are treated
with the use of duality and of the non-renormalization theorems
for the central charges of the SUSY algebra. A combination of these two ideas
allows one to predict the spectrum of the stable BPS states which become
massless
in the deep non-perturbative region.

\item
{Duality}.

The duality transformation can be easily defined
in the finite ${\cal N}=4$ SUSY theory just as the modular transformations
generated by $\tau \to \tau^{-1}$ and $\tau \to \tau +1$.The complexified 
coupling constant  
plays the role of modulus of the auxiliary elliptic curve.
This makes a strong hint that
the duality can be related with a modular space of some Riemann surfaces,
where the modular group acts.

\item {Non-perturbative RG flows and auxiliary Riemann surface}

In the asymptotically free
theory, one has to match the duality with the renormalization
group. This is non-trivial,
since now
$\tau$ depends on the scale which is  involved into the duality
transformation.  The duality acts on the moduli space
of vacua and this moduli space is associated with the moduli space of
the auxiliary Riemann surface, where the modular transformations act.

At the next step, one has to find out proper variables whose modular
properties fit the field theory interpretation. These variables are
the integrals of a meromorphic 1-form $dS$
over the cycles on the Riemann surface,
$a_i$ and $a_{D,i}$

\be\label{aad}
a_{i}=\oint_{A_{i}}dS, \qquad
a_{D,i}=\oint_{B_{i}}dS,
\ee
(where $i,j=1,....,N-1$ for the gauge group $SU(N)$).

These integrals play the two-fold role in the Seiberg-Witten approach.
First of all, one may calculate the prepotential $\cal{ F}$ and, therefore,
the low
energy effective action through the identification of
$a_D$ and $\frac{\partial \cal{F}}{\partial a}$ with
$a$ defined as a function of moduli  by formula
(\ref{aad}).
Then, using the property of the differential $dS$
that its variations w.r.t. moduli are holomorphic one may also calculate the
matrix of coupling constants

\be\label{Tij}
T_{ij}(u)=\frac{\partial^{2}{\cal{F}}}{\partial a_{i} \partial a_{j}},
\ee

The second role of formula (\ref{aad}) is that, as
was shown in \cite{SW1,SW2} these integrals define the spectrum of the stable states
in the theory which saturate the Bogomolny-Prasad-Sommerfeld (BPS)
limit. For instance, the formula for the BPS spectrum in the $SU(2)$
theory reads as
\be\label{BPS}
M_{n,m}(u)=\left|na(u) +ma_{D}(u)\right|,
\ee
where the quantum numbers $n,m$ correspond to the ``electric" and ``magnetic"
states. 

The column $(a_i,a_{D,i})$ transforms under
the action of the modular group  as a section of the linear bundle.
Its global behavior, in particular, the structure of the singularities
is uniquely determined by the monodromy data. As we discussed earlier,
the duality transformation connects different singular points. In particular,
it interchanges ``electric", $a_i$ and ``magnetic", $a_{D,i}$
variables which describe the perturbative degrees of freedom at the
strong and weak coupling regimes 
respectively.
Manifest calculations involving the Riemann surface allow one to analyze
the monodromy properties of dual variables when
moving in the space of the order parameters. For instance,
in the simplest $SU(2)$ case, on the $u$-plane of the single order parameter
there are three singular points, where some cycle shrinks and the magnetic and
electric variables mix when encircling these points.
\end{itemize}

Summarizing, the information about the Seiberg-Witten solution is encoded
in the Riemann surface bundled over the moduli space of vacua and the
meromorphic differentials on this Riemann surface encoding the spectrum
of BPS states and prepotential. It was observed in \cite{gkmmm} that these
data are standard ingredients of the integrable holomorphic many-body systems.

\section{SW solution and classical holomorphic systems}
The complexified periodic Toda chain is a one-dimensional system of $N$ non-relativistic particles interacting with the following potential
$$
V(x_1, ... , x_N) = \Lambda^2 \sum_{i=1}^{N} e^{x_{i}-x_{i+1}}, \,\,\,\,\,\,\, x_{N+1} = x_{1}
$$
where the coordinates $x_i \in \mathbf{C}/(2 \pi \mathbf{Z})$ while the momenta $p_{i} \in \mathbf{C}$.
The Riemann surface and meromorphic differentials involved
into the solution for pure $SU(N)$ SYM theory are intrinsic 
objects for the N-body periodic Toda chain \cite{gkmmm,MW,6} whose equations of motion read

\be\label{Todaeq}
\frac{\partial q_i}{\partial t} = p_i \ \ \ \ \
\frac{\partial p_i}{\partial t} = e^{q_{i+1} -q_i}-
e^{q_i-q_{i-1}}
\ee
where the periodic boundary conditions $p_{i+N}=p_i$, $q_{i+N}=q_i$ imposed.
In the Toda system there are exactly $N$ conservation laws
constructed from the Lax operator. 
The Lax operator is represented by the $N\times N$
matrix depending on dynamical variables

\be\label{LaxTC}
{\cal L}^{TC}(w) =
\left(\begin{array}{ccccc}
p_1 & e^{{1\over 2}(q_2-q_1)} & 0 & & we^{{1\over 2}(q_1-q_{N})}\\
e^{{1\over 2}(q_2-q_1)} & p_2 & e^{{1\over 2}(q_3 - q_2)} & \ldots & 0\\
0 & e^{{1\over 2}(q_3-q_2)} & -p_3 & & 0 \\
 & & \ldots & & \\
\frac{1}{w}e^{{1\over 2}(q_1-q_{N})} & 0 & 0 & & p_{N}
\end{array} \right)
\ee
The characteristic equation for the Lax matrix

\be\label{SpeC}
{\cal P}(\lambda,w) = \det_{N\times
N}\left({\cal L}^{TC}(w) - \lambda\right) = 0
\ee
generates the conservation laws and
determines the spectral curve.
If one restores the dependence
on $\Lambda_{QCD}$ in this spectral curve, it takes the form
\be\label{fsc-Toda'}
w + \frac{\Lambda_{QCD}^{2N}}{w} = 2P_{N}(\lambda )
\ee
where $P_N(\lambda)= \det(\phi -\lambda)$,$\phi = diag(a_1,\dots,a_N)$.
The spectral curve exactly coincides with the Riemann surface
introduced in the context of $SU(N)$ gauge theory. It can be also
put into the hyperelliptic form
\be\label{2}
2Y\equiv w-{1\over w},\ \ \ Y^2=P_{N}^2(\lambda)-1
\ee
The integrals of motion parameterize the moduli space of the
complex structures of the hyperelliptic surfaces of genus $N-1$, which is
the moduli space of vacua in physical theory.

After having constructed the Riemann surface and the moduli space
describing the $4d$ pure gauge theory,
we turn to the third crucial ingredient of solution
that comes from integrable systems, the generating differential $dS$.
This differential is
in essence the ``shorten" action $pdq$. Indeed, in order to construct
action variables, $a_i$
one needs to integrate the differential ${\tilde {dS}}=\sum_ip_idq_i$
over $N-1$ non-contractable cycles in the Liouville torus which is nothing
but
the level submanifold of the phase space, i.e. the submanifold defined by
values of all $N-1$ integrals of motion fixed. On this submanifold, the
momenta $p_i$ are functions of the
coordinates, $q_i$. The Liouville torus 
is just the Jacobian
corresponding to the spectral curve.

In the general case of a $g$-parameter family of complex curves
of genus $g$, the Seiberg-Witten differential
$dS$ is characterized by the property
$\delta dS = \sum_{i=1}^g \delta u_i dv_i$,
where $dv_i(z)$ are $g$ holomorphic 1-differentials
on the curves , while $\delta u_i$ are
variations of $g$ moduli along the base.
In the associated integrable system, $u_i$ are integrals of motion
and $\pi_i$, some $g$ points on the curve are momenta.
The symplectic structure is
\be\label{dSjac}
\sum_{i=1}^g da_i\wedge dp_i^{Jac} = \sum_{i,k=1}^g
du_i\wedge dv_i(\pi_k)
\ee
The vector of the angle variables,
\be\label{pjac}
p_i^{Jac} = \sum_{k=1}^g \int^{\pi_k} d\omega_i
\ee
is a point of the Jacobian, and the Jacobi map identifies this with the
$g$-th power of the curve, $Jac\ \cong {\cal C}^{\otimes g}$.
Here $d\omega_i$ are {\it canonical} holomorphic differentials,
$dv_i = \sum_{j=1}^g d\omega_j \oint_{A_j} dv_i$.

In $SU(2)$  case, the spectral curve is
\be\label{sc2}
w+{1\over w}=2(\lambda^2-u)
\ee
while $u=p^2-\cosh q$. Therefore, one can write for the action variable
\be\label{dSTC}
a=\oint pdq=\oint \sqrt{u-\cosh q}dq=\oint \lambda {dw\over w},\ \ \
dS=\lambda {dw\over w}
\ee
where we made the change of variable $w=e^q$ and used equation (\ref{sc2}).

One can easily check that the derivatives of $dS$ with respect to moduli are
holomorphic, up to total derivatives. Say, if one
parameterizes
$P_{N}(\lambda)=-\lambda^{N}+s_{N-2}\lambda^{N_c-2}+...$ and note
that $dS=\lambda dw/w=\lambda dP_{N}(\lambda)/Y=P_{N}(\lambda) d\lambda/Y
+ \hbox{ total derivatives }$, then
\be
{\partial dS\over\partial s_k}={\lambda^kd\lambda\over Y}
\ee
and these differentials are holomorphic if $k\le N-2$.
Thus, $N-1$ moduli gives rise to $N-1$
holomorphic differentials which perfectly fits the genus of the curve (we
use here that there is no modulus $s_{N-1}$).

If more complicated SYM theory with different matter content is considered
the periodic Toda chain gets substituted by more general Hamiltonian system
but the scheme is generic. The spectral curve in each case gets
identified with the SW curve while the "$pdq$" differential gets 
identified with the SW differential. At the classical level this
relation  looks like a curious coincidence however
later the deep reason behind the correspondence between 
the integrable systems and SYM theories has been recognized.

\section{Nekrasov partition function and quantization of integrable system}

\subsection{Instanton partition function}

First, let us recall the definition of the Nekrasov partition function \cite{nekrasov, NO02} of a four dimensional gauge theory in the $\Omega$-background $\mathbf{C}^2_{\ep, \hbar}$ where $\ep, \hbar$ are two equivariant parameters for the torus action
on the instanton moduli space.
In this Section we adopt  $\epsilon_1=\epsilon, \epsilon_2=\hbar$  to fit with QM notation. 
The $\Omega$-background was initially introduced to regularize the integrals over the instanton moduli space however it turned out that it provides a lot of additional information about the instanton ensembles and serves
as crucial ingredient for the explanation of the origin of the integrable systems.

Let $\mathbf{a}$ denote a set of complex scalars which parametrize the moduli space of vacua, $\mathbf{m}$ is set of masses for fundamental matter multiplets and $\Lambda^{2N} = \exp (2 \pi i \tau)$ is a generated mass scale that counts instantons. 
The full Nekrasov partition function consists of perturbative and non-perturbative contributions
\beq
Z(\mathbf{a},\mathbf{m},\Lambda; \ep, \hbar) = Z^{pert.}(\mathbf{a},\mathbf{m},\Lambda; \ep, \hbar) \times Z^{inst.}(\mathbf{a},\mathbf{m},\Lambda; \ep, \hbar)
\eeq
The non-perturbative part of the partition function is obtained by the equivariant localization on the instanton moduli space and is defined as follows. Let
\beq
\mathcal{V}_{\lambda} = \sum_{i=1}^{N} \sum_{(r,s) \in \lambda_{i}} e^{a_i + (r-1)\ep + (s-1) \hbar}, \,\,\,\,\,\,\, \mathcal{W} = \sum_{i=1}^{N} e^{a_i}, \,\,\,\,\, \mathcal{M} = \sum_{i=1}^{N_f} e^{m_i}
\eeq
\beq
\mathcal{T}_{\lambda} = - \mathcal{M} \mathcal{V}_{\lambda}^* + \mathcal{W} \mathcal{V}_{\lambda}^* + e^{\ep+\hbar} \mathcal{V}_{\lambda} \mathcal{W}^* - (1 - e^{\ep}) (1-e^{\hbar}) \mathcal{V}_{\lambda} \mathcal{V}_{\lambda}^*
\eeq
which appear as the characters of the natural bundles on the instanton moduli space at a fixed point, parametrized by a set $\{ \lambda_i \}$ of $N$ Young diagrams. 
The collection of fixed points is defined as follows. Consider the contribution from the instanton charge K 
and introduce partition $K=\sum_{1}^{N}k_i$ where $k_i\ge 0$. Introduce $\lambda_{l}$ - the partition of $k_l$
and consider the distribution of K boxes among the collection of N Young tableau $\vec{\lambda}=(\lambda_1,\dots,\lambda_N)$.
We denote $|\lambda_l|=k_l$ the number of boxes in l'th diagram hence $\sum_{l} |\lambda_l|=K$. The fixed points
from the l'th Young tableau which contribute the partition function are located at   
\beq
\phi_l= a_l + \epsilon(r-1) +\hbar(s-1)
\eeq
where the pair $(r,s)$ corresponds to the box in the l'th tableau.

The star-operation inverts the weights of all character: 
\beq
\biggl(\sum_{a} e^{w_a} \biggr)^* =  \sum_{a} e^{-w_a}
\eeq
The instanton partition function can be written as
\beq
Z^{inst.} (\mathbf{a} ,\mathbf{m}, \Lambda; \ep, \hbar) = \sum_{\{\lambda\}} \Lambda^{2N |\lambda|} e(\mathcal{T}_{\lambda})
\eeq
where
\beq
e(\sum_{a}e^{w_a}-\sum_b e^{w_b}) = \frac{\prod_b w_b}{\prod_a w_a}
\eeq
is a symbol that converts the sum of characters into the product of weights. 

The perturbative part is more subtle due to ambiguity of the boundary conditions at infinity \cite{nikitakz}. It can be written as
\beq
Z^{pert.}(\mathbf{a},\mathbf{m},\Lambda;\ep,\hbar) = \Lambda^{-\frac{N}{\ep \hbar} \sum_{i=1}^{N} a_i^2} e\biggl( \frac{e^{\ep+\hbar}(\mathcal{M} \mathcal{W}^* -\mathcal{W} \mathcal{W}^*)}{(1-e^{\ep})(1-e^{\hbar}) }\biggr)
\eeq 
but since the character now has infinitely many terms a proper regularization is needed. Physically the first multiplier comes from the tree level contribution, while the first and the second terms in the character come from the one-loop contribution of the matter multiplets and $W$-bosons, respectively.

\subsection{Nekrasov-Shatashvili limit}

The low-energy description of undeformed four dimensional gauge theory is characterized by the prepotential, which can be obtained as the limit of the deformed partition function
\beq
\mathcal{F}(\mathbf{a}, \mathbf{m}, \Lambda) =  \lim_{\ep,\hbar \to 0} \ep \, \hbar \log Z(\mathbf{a},\mathbf{m}, \Lambda; \ep,\hbar)
\eeq

In \cite{NS3} a refinement of the correspondence with integrable systems was proposed. It was argued that the effective two dimensional theory, which appears in the limit $\ep \to 0$, is related to a corresponding quantized algebraic integrable system, and $\hbar$ plays the role of the quantization parameter. 

More precisely we can consider a four dimensional gauge theory on $\mathbf{C} \times \mathbf{D}_{\hbar}$ where $\mathbf{D}_{\hbar}$ is the cigar-like geometry of \cite{nekwit} with the $\Omega$-deformation turned on. With an appropriate twist this geometry breaks half of the supersymmetries. Upon choosing the boundary conditions on $\mathbf{C} \times \p \mathbf{D}_{\hbar} = \mathbf{C} \times S^{1}$ which preserve the remaining supersymmetries, we can reduce our theory to a two-dimensional $\mathcal{N}=(2,2)$ theory on $\mathbf{C}$, the low energy description of which is characterized by the effective twisted superpotential
\beq
\mathcal{W}(\mathbf{a}, \mathbf{m}, \Lambda; \hbar) = \lim_{\ep \to 0} \ep \log{Z(\mathbf{a}, \mathbf{m}, \Lambda; \hbar)}
\eeq
For the perturbative contribution we have
\beq
\mathcal{W}^{pert.}(\mathbf{a}, \mathbf{m}, \Lambda; \hbar) = \lim_{\ep \to 0} \ep \log Z^{pert.}(\mathbf{a},\mathbf{m},\Lambda;\ep,\hbar)  =  
\eeq
$$
-\frac{1}{2 \hbar} \log \biggl( \frac{\Lambda^{2N}}{\hbar^{2N}} \biggr) \sum_{i=1}^{N} a_i^2 + \sum_{i,j=1}^{N} \varpi_{\hbar}(a_i - a_j) - \sum_{i=1}^{N}\sum_{a=1}^{N_f}  \varpi_{\hbar}(a_i - m_a)
$$
where $\varpi_{\hbar}(x)$ obeys
\beq
\label{varpi}
\frac{d}{dx} \varpi_{\hbar} (x) = \log \Gamma \biggl( 1 + \frac{x}{\hbar} \biggr) = const. - \sum_{n=1}^{\infty} \log \biggl(\frac{x+n \hbar}{\hbar} \biggr)
\eeq

The one-loop contribution $\varpi_{\hbar}(m)$ has a simple intuitive explanation as a contribution of infinite number of angular momentum modes with mass parameters $(m+n \hbar)$ for $n$-th mode into the effective twisted superpotential, which are chiral multiplets in the effective two dimensional theory. Indeed, after integrating out a single chiral multiplet we get
\beq
\Delta \mathcal{W}_n = -(m+n\hbar) \biggl[\log\biggl(\frac{m+n \hbar}{\hbar}\biggl)-1\biggr]
\eeq
and after summing over $n$ we get $\varpi_{\hbar}(m)$.

There are two natural boundary conditions \cite{nekwit} which require
\beq
Type A: \,\,\,\,\,\,\,\,\, \frac{a^{D}_i}{\hbar} = \frac{\p \mathcal{W}(\mathbf{a},\mathbf{m},\Lambda; \hbar)}{\p a_i} \in  \mathbf{Z}
\eeq
\beq
Type B: \,\,\,\,\,\,\,\,  \frac{a_i}{\hbar} \in \mathbf{Z} + \frac{\theta_i}{2 \pi}, \,\,\,\,   \theta \in [0,2 \pi)
\eeq
Type $A$ corresponds to Neumann-type boundary condition for gauge fields leading to a dynamical vector multiplet in two dimensions. On the contrary type $B$ corresponds to Dirichlet boundary conditions, fixing the holonomy along the boundary of the cigar parametrized by $\theta_i$ and freezing gauge degrees of freedom. The choice of this boundary conditions specifies the quantization of an algebraic integrable system. 

The Bethe/gauge correspondence states that the vacua of the effective two dimensional theory are in one-to-one correspondence with the eigenstates of the Hamiltonians of the quantum integrable system. Moreover, the expectation values of the topologically protected chiral observables, which are traces of the adjoint scalars in a vector multiplet $Tr \mathbf{\phi}^k$ of a four dimensional theory, coincide with the eigenvalues of the Hamiltonians $\mathcal{H}_{k}$ on this states:
\beq
\langle vac | \mathcal{H}_{k} | vac \rangle
 \longleftrightarrow \langle Tr \mathbf{\phi}^k \rangle_{vac.}  
\eeq
We are interested in the quantization of this Hamiltonian system. The Hamiltonians and the momenta are now promoted to differential operators, acting on wave functions $\psi(x_1,..., x_N)$ with $p_i = \hbar \p_{i}$. There are two natural quantizations, corresponding to type $A$ and type $B$ boundary conditions \cite{NS3}. Type $A$ quantization corresponds to $x_i \in \mathbf{R}$ and $\psi(x_1- \bar{x}, ..., x_N - \bar{x}) \in L^2(\mathbf{R}^{N-1})$ where $\bar{x} = \sum_{i}^{N} x_i/N$ is the center of mass mode, which decouples in a trivial way. This condition leads to discrete unambiguous spectrum that corresponds to the set of vacua in the gauge theory, provided that the type $A$ boundary condition $a_{D}/\hbar \in \mathbf{Z}$ is satisfied. 

Type $B$ quantization corresponds to $x_i \in i \mathbf{R}/2 \pi \mathbf{Z}$ and quasi-periodic non-singular wave functions 
\beq
\psi(x_1, ... , x_a + 2 \pi i, ... , x_N) = e^{i \theta_a} \psi (x_1, ..., x_N)
\eeq
The quasi-periodicity parameters $\theta_{a}\in [0,2\pi)$ are also known as Bloch-phases. In the special case of $N=2$, after the decoupling of the center of mass mode, the equation on the wave function coincided with Mathieu equation
\beq
-\hbar^2 \psi''(x) + 8 \Lambda^2 \cos(2 x) \psi(x) = 8 u \psi(x)
\eeq
where $u = \frac14 \langle Tr \mathbf{\phi}^2 \rangle$. For real $\Lambda$ and $\hbar$ it describes a particle moving in a periodic cosine potential. At fixed $\theta$-parameters the spectrum is discrete. However as we vary them the spectrum consists of peculiar structure of bands and gaps. In particular for small $\Lambda$ and when we sit near the edge of some band, the spectrum has exponentially small in $\sim 1/\hbar$  splitting of the eigenvalues. More precisely if $\theta_1 = ... = \theta_{k_1}; \theta_{k_1+1} = ... = \theta_{k_2}; ... ; \theta_{k_m+1} = ... = \theta_{k_N}$ and are all integers, then instead of naive $\frac{N!}{k_1!k_2!...k_N!}$-fold degeneracy as for $\Lambda=0$ we have non-degenerate spectrum due to quantum reflection on the potential.

This second type of quantization appears to be more mysterious from the gauge theory point of view. When all $\theta$-parameters are zero $a_a/\hbar$ are forced to be equal to integer, and when $a_{ab}/\hbar \in \mathbf{Z} \setminus \{0\}$ some perturbative $W$-boson modes naively become massless that is clear from the logarithmic singularities in their perturbative contribution into the effective twisted superpotential. Recently new interesting aspects of Bethe/gauge correspondence were elaborated in \cite{bullmore}.

Let us emphasize that the Bethe/gauge correspondence yields the connection between the Nekrasov partition
function in the NS limit and the ingredients of the quantum integrable systems. However it does not tell
what are degrees of freedom in this system. To answer this question we have to add the second
crucial ingredient - AGT correspondence \cite{agt,wyllard}
 
\section{AGT correspondence}

Let us briefly remind that according the AGT correspondence the Nekrasov partition function
for $SU(2)$ is identified with the particular conformal block in the Liouville theory whose
type depends on the matter content on the gauge theory side. 
The central charge $c$ in the Liouville theory
is expressed in terms of the parameters of the $\Omega$-deformation 
as follows
\beq
c= 1+6Q^2,\qquad Q=b+ \frac{1}{b},\qquad b^2=\frac{\epsilon_2}{\epsilon_1}
\eeq
hence the NS limit $\epsilon_2 \rightarrow 0$ on the gauge theory side corresponds to the classical $c \rightarrow \infty$ limit in the
Liouville theory. The dimensions of the degenerate operators in the classical limit behave as 
\beq
h_{s,1}= -\frac{s-1}{2} + O(1/c) \qquad h_{1,r}= -\frac{r^2-1}{24}c + O(c^0)
\eeq

The operators are naturally classified at large $c$ limit according to  behavior of
their conformal dimensions $\Delta_i$.
The operators whose dimensions are proportional to $c$ are called heavy while ones whose
dimensions are $O(1)$ are called light operators. It is natural to introduce the
classical dimensions $\delta_i$ for the heavy operators defined as 
\beq
\Delta_i= b^{-2}\delta_i
\eeq
According to AGT correspondence the 4-point  spherical Liouville conformal block 
is identified with the total Nekrasov partition function for $SU(2)$ $N_f=4$ theory
\beq
\mathcal{F}_{a}(\epsilon_1,\epsilon_2,m_i, q) = Z_{tot}(a,\epsilon_1,\epsilon_2,m_i,\tau)
\eeq
\beq
Z_{tot}=Z_{cl}Z_{1-loop}Z_{inst}
\eeq
where the factors correspond to the classical, one-loop and instanton contribution
to the partition function.
The coordinate at the Coulomb branch $a$ corresponds to the intermediate conformal dimension 
in the conformal block, masses $m_i$  yield the corresponding conformal dimensions 
of insertions $\delta_i$ and the complexified coupling constant $\tau $
in SYM theory gets mapped
into the conformal cross-ratio in the 4-point spherical conformal block \cite{agt}.
The 4-point correlator in the Liouville theory is expressed in terms the
Nekrasov partition function as follows \cite{agt}
\beq
\lal V(0)V(\infty)V(1)V(q) \rr \propto \int da a^2|Z_{tot}(a)|^2
\eeq

In the classical NS limit
$\epsilon_2\rightarrow 0$ the twisted superpotential gets identified with the
classical conformal block .
\beq
Z_{inst}(a,\epsilon_1,\epsilon_2,\tau)\rightarrow \exp(b^{-2} W(a,m_i,\hbar,\tau))
\eeq
\beq
\mathcal{F}_{a}(\epsilon_1,\epsilon_2,\delta_i, q) \rightarrow \exp(b^{-2} f_{\delta_{in}}(\delta_i,\hbar,q)),\qquad
\delta_{in}= b^{-2}(\frac{1}{4} - \frac{a^2}{\hbar^2})
\eeq
Since we have exact coincidence of the twisted superpotential and the classical conformal block
the naive poles in the twisted superpotential correspond to the naive poles
in the intermediate dimension plane  in the classical 
Liouville conformal block. 

Similarly the torus one-point classical conformal block corresponds 
to twisted superpotential in $N=2^*$ theory which depends
on the external and intermediate  dimensions $\Delta,\Delta_{in}$.
To have the unified picture it is useful to represent the one-point torus conformal block $\mathcal{F}_{c,\Delta_{in}}(\lambda, q)$ as the  spherical
conformal block with four insertions $\mathcal{F}_{c,\Delta_{in}}^{sp}[\lambda_1,\lambda_2,\lambda_3,\lambda_4](q)$.
The explicit mapping of parameters under the map goes in arbitrary $\Omega$-background as follows
\beq
\mathcal{F}_{c,\Delta_{in}}(\lambda,q) = \mathcal{F}_{c,\Delta_{in}}^{sp}[\frac{\lambda}{2} - \frac{1}{2b},
\frac{\lambda}{2} + \frac{1}{2b},\frac{b}{2}, \frac{b}{2}](q)
\eeq
where conformal dimensions $\Delta_i$ are equal to
\beq
\Delta_i=\frac{1}{4}(Q^2 -\lambda_i^2) 
\eeq
The  modulus of the torus $q$ gets mapped into the position of 
the insertion $x$ on the sphere as 
\beq
q(x)= \exp \l -\frac{\pi K(1-x)}{K(x)} \r \qquad K(x)= \int_0^1 \frac{dt}{\sqrt{(1-t^2)(1-xt^2)}}
\eeq

At the classical $b\rightarrow 0$ limit the corresponding classical 4-point spherical conformal
block  representation 
\beq
f^{sp}_{\delta_{in}}[\delta_1,\delta_2,\delta_3,\delta_4](q) = f^{sp}_{\frac{1}{4} - \frac{a^2}{\hbar}}[\frac{1}{4}(1 -\frac{m^2}{\hbar}),\frac{1}{4}(1 -\frac{(m+\hbar)^2}{\hbar}),
\frac{1}{4},\frac{1}{4}](q) 
\eeq
where $m$ is the adjoint mass in $N=2^*$ theory.
 All of insertions  correspond to the heavy operators and
the intermediate classical dimension is heavy as well. 

If we insert the additional light $\Psi_{2,1}(z)$ operator in 1-point
torus conformal block and consider the 2-point classical conformal
block the Lame equation can be identified \cite{fateev}.
The decoupling equation for the 2-point block can be brought into the
conventional QM form with Lame potential
\beq
(\frac{d^2}{dz^2}+ k \wp(z|\tau) +E)\Psi(z,\tau,E)=0
\eeq
where $\tau$ is the modulus of the elliptic curve and 
$k= \frac{m}{\hbar}(\frac{m}{\hbar} +1)$
The energy in the $\wp$ potential is related with the classical 
conformal block via the properly normalized quantum Matone relation \cite{poghossian}
\beq
\frac{E}{4\pi^2}= -\l \frac{a}{\hbar} \r^2 -\frac{k}{12}(1 - 2E_2(\tau))
+\hbar^{-1}q\frac{d}{dq}W^{N=2^*}(q,a,m,\hbar) 
\eeq
where $E_2$ is the Eisenstein series and $a$ is  the Bloch phase.

To get the Mathieu equation one has to take the Inozemtsev limit
in the decoupling equation for the degenerate irregular conformal block. The 
wave function of two-particle Toda system is represented as 
the matrix element of the degenerate chiral vertex operator between two Gaiotto 
states \cite{gaiotto}. 
\beq
\lal \Delta_1,\Lambda^2|V_{+}(z)|\Delta_2,\Lambda^2 \rr \rightarrow z^{\Delta_1 -\Delta_{+} -\Delta_2}
\phi(\frac{\Lambda}{\hbar},z) \exp \l \frac{1}{b^2} f_{\delta}(\frac{\Lambda}{\hbar}) \r
\eeq
The function 
\beq
\psi( \frac{\Lambda}{\hbar},z) = z^{r}\phi(\frac{\Lambda}{\hbar},z), \qquad \delta= 1/4 - r^2
\eeq
obeys the Mathieu equation with the energy 
\beq
E= 4r^2 -\Lambda \frac{\partial}{\partial \Lambda}f_{\delta}(\frac{\Lambda}{\hbar})
\eeq
Given the solutions to the decoupling equations we can evaluate the 
monodromy matrix. The Zamolodchikov's method \cite{zamo} allows to extract 
the conformal block itself from the monodromy data.

Concluding this Section note that the QM enjoys several exponentially small 
non-perturbative phenomena
like the energy splitting or the small gaps well above the barrier. The natural
question concerns the interpretation of these phenomena at the SYM side. The 
whole picture has not been developed yet however a few results have to be mentioned.
It was demonstrated in \cite{gmn} that to reproduce the gaps above the barrier one
has to perform the peculiar resummation of the non-perturbative contributions
first applied in this context in \cite{beccaria}. Another approach based of the
summing over the complex saddles has been suggested in \cite{nikitanew}.

\section{RG flows and Whitham dynamics}

The prepotential which sums up the perturbative and instanton contributions
depends on the vacuum moduli $\vec{a}$ and the parameters of the 
the Lagrangian. For the finite SYM theory the useful parameter is the 
bare coupling constant $\tau$ while in the asymptotically free theories
the corresponding parameter is the $log \Lambda$. The dependence of the prepotential on these
parameters is covered by the non-perturbative renormalization group (RG) dynamics
which fits with the integrable
systems in the interesting way.

It turned out \cite{gkmmm,gmmm,Jose1,6, KDP1} that on the top of the holomorphic integrable
system like closed Toda chain and elliptic Calogero system there is the second
Hamiltonian system which is  example of the Whitham Hamiltonian system. Such systems
are related to the perturbation of the finite-gap solutions parametrized by the
Riemann surfaces. The phase space of the Whitham system relevant fro SW solution is finite-dimensional
and is equipped with by the Poisson bracket
\beq
\{a_i,a_{D,j}\}= \delta_{ij}
\eeq
The time variable for the Toda chain is identified with the $t=log \Lambda$
and the Hamiltonian gets identified with $H=<vac|Tr\phi^2| vac>$.  
The Whitham equations read as 
\beq
a_{D,i}= \frac{\partial \cal{F}}{\partial a_i}\qquad  E= -\frac{\partial \cal{F}}{log \Lambda}
\eeq
therefore the prepotential plays the role as the action $S(x)$ for the Hamiltonian dynamics.
Let us emphasize that the Hamilton-Jacobi equation in the Whitham system plays 
the role of the anomaly equation at the gauge theory side. Note that the Hamiltonian
equation in the Whitham dynamics
\beq
\frac{da_{D,i}}{dlog \Lambda} = \frac{dH}{da_i}
\eeq
plays an interesting meaning in the first Hamiltonian system like Toda chain \cite{gm}.
It gets identified with the so-called P/NP relation \cite{pnp1,pnp2} which connects the 
perturbative and non-perturbative contributions in the 
corresponding quantum mechanical system.

In the $\Omega$-deformed case in the Nekrasov-Shatashvili limit the Whitham
system remains to be classical and only Hamiltonians get deformed \cite{ poghossian}.
If both deformation parameters are included the Whitham dynamics gets quantized
\cite{nikitakz} and the product $\epsilon_1 \epsilon_2$ plays the role of the
Planck constant for the Whitham dynamics. Since the prepotential according
to AGT correspondence gets identified with the conformal block in Liouville and Toda
theory the Whitham equation coincides with the KZ equation for the dependence
of conformal block on the moduli space of the complex structures of the punctured
Riemann surfaces.

It it possible to derive the Whitham hierarchy for the SYM theory perturbed
by the terms $T_kTr \Phi^k$ when $T_k$ plays the role of higher times  
in the Whitham hierarchy and the prepotential plays the role of the
semi-classical tau-function \cite{gkmmm, gmmm, Jose1, KDP1}. The prepotential in the theory 
without the $\Omega$- deformation obeys the WDVV equations \cite{wdvv}.

\section{Branes as degrees of freedom in integrable systems}

Let us briefly describe the geometrical aspects of the correspondence. The key 
players in the integrable dynamics  behind SYM are D-branes which carry the
abelian gauge field on their worldvolume \cite{polchinski1}. The gauge theories
with the prescribed  gauge group matter content can be manufactured from the set of 
branes of different dimensions \cite{witten1} in particular the SW solution can be
derived via M5 brane wrapped around the particular Riemann surface \cite{witten2,vafa}.
The set of T and S dualities allows to transform one brane picture into another.
Generically the coordinates of the branes in the directions  transverse
to their world-volume in 10d play the role of the vacuum moduli or coupling constants.
The back reaction of one brane into another due to the non-vanishing tensions
amounts to their nontrivial profiles which corresponds to the particular 
solution to the RG equations. 

It is not surprise that integrable systems of Toda and Calogero type are 
intimately related with the brane configurations. The clear cut reason 
behind this relation is
as follows. It was found in \cite{gn} that the elliptic Calogero model
is the particular example of the Hitchin system \cite{hitchin} on the 
cotangent bundle to the moduli space of the holomorphic G-valued 
bundles over the Riemann surface $\Sigma$ with n marked points $T^{*}M_{(G, \Sigma,n)}$.
The elliptic Calogero system  corresponds to the genus one surface 
with one marked point. The Hitchin system as the system on the D-branes
has been described in \cite{dw} and the separation of variables 
in the Hitchin system which allows to represent the phase space in terms of D0 branes as 
Hilbert space of points on $T^*\Sigma$ has been found in \cite{gnr}.

In spite of this well justified relation between Calogero and Toda system
and  theories on the D-brane worldvolumes it took some time to understand
what type of branes  should be considered in SYM to explain
the hidden integrability. It was argued in \cite{gorsky,ggm1} that integrability
deals with the artificially added defect probe branes however the type
of these branes has not been clarified. The answer to this question
has been found when the AGT correspondence has been formulated.
It turned out that in the Liouville or Toda theory  one has to add the particular surface  operator
introduced in \cite{gw}. 

In the Liouville and Toda theory this defect corresponds
to the operator which is degenerate at the second level $\Psi_{2,1}$ \cite{gukov}
and therefore obeys the second order differential equation which we have 
discussed in the previous Section. It is this equation
which plays the role of Schrodinger equation for the integrable models.
For instance the decoupling equation for the two-point function on the torus 
in the Liouville theory obeys the Schrodinger equation for the elliptic
Calogero model \cite{fateev}. One can recognize the interplay between
the surface operators and the integrable systems at the classical level
as well \cite{ggs}.

Hence the steps are as follows: first one adds the additional brane - degenerate operator, investigate the 
monodromy properties of the solution to the decoupling equation and extract from
monodromy the expression for the conformal block without the insertion 
according to the recipe \cite{zamo}.
There is some analogue of this procedure in the superconductivity. One can
consider the one-particle excitation at the top of the condensate. This
excitation does not ruin the condensate ground state and investigating 
its properties one can extract the important information
concerning its ground state.

Summarizing we could claim that the degrees of freedom in the finite-dimensional 
system are  coordinates of particular branes in the internal space. A
dynamics itself reflects the consistency of the whole brane configuration
with the non-perturbative RG flows while quantization is provided by
the AGT correspondence for the Nekrasov partition function.

I would like to thank all my colleagues with whom I shared the pleasure to collaborate
and discuss these interesting issues.The work was performed at the Institute for Information 
Transmission Problems with the financial support of the Russian Science Foundation (Grant No.14-50-00150)

\end{document}